\documentclass[final]{aipproc}

\layoutstyle{6x9}

\usepackage{bm}
\usepackage{slashed}
\usepackage{graphicx}
\usepackage{subfigure}
\makeatletter
\newcommand\setcaptype[1]{\def\@captype{#1}}
\makeatother

\usepackage[svgnames]{xcolor}
\usepackage[normalem]{ulem}

\begin{document}

\title{Monte-Carlo Approach to Calculating the Fragmentation Functions in NJL-Jet Model}


\author{Hrayr H.~Matevosyan}{
  address={CSSM, School of Chemistry and Physics, \\
University of Adelaide, Adelaide SA 5005, Australia\\
http://www.physics.adelaide.edu.au/cssm
}
}

\author{Anthony W. Thomas}{
  address={CSSM, School of Chemistry and Physics, \\
University of Adelaide, Adelaide SA 5005, Australia\\
http://www.physics.adelaide.edu.au/cssm
}
}

\author{Wolfgang Bentz}{
  address={Department of Physics, School of Science,\\  Tokai University, Hiratsuka-shi, Kanagawa 259-1292, Japan\\
http://www.sp.u-tokai.ac.jp/} 
}

\begin{abstract}
Recent studies of the fragmentation functions using the Nambu--Jona-Lasinio (NJL) - Jet model have been successful in describing the quark fragmentation functions to pions and kaons. The NJL-Jet model employs the integral equation approach to solve for the fragmentation functions in quark-cascade description of the hadron emission process, where one assumes that the initial quark has infinite momentum and emits an infinite number of hadrons. Here we introduce a Monte Carlo (MC) simulation method to solve for the fragmentation functions,, that allows us to relax the above mentioned approximations. We demonstrate that the results of MC simulations closely reproduce the solutions of the integral equations in the limit where a large number of hadrons are emitted in the quark cascade. The MC approach provides a strong foundation for the further development of the NJL-Jet model that might include many more hadronic emission channels with decays of the possible produced resonances, as well as inclusion of the transverse momentum dependence (TMD), all of which are of considerable importance to the experimental studies of the transverse structure of hadrons.
\end{abstract}

\maketitle


\section{Introduction}
\label{NJL-JET} 

 The novel efforts to extract the quark fragmentation functions from various experimental data \cite{Hirai:2007cx,deFlorian:2007aj} have generated a renewed interest in the long studied subject of analyzing hard scattering reactions  \cite{Field:1977fa,Altarelli:1979kv,Collins:1981uw,Jaffe:1996zw,Ellis:1991qj,Barone:2001sp,Martin:2003sk}. The analysis of the transversity quark distribution functions \cite{Barone:2001sp,Ralston:1979ys} and a variety of other semi-inclusive processes \cite{Sivers:1989cc,Boer:2003cm} also critically depend on the knowledge of the fragmentation functions.
  
\begin{figure*}[ptb]
\centering 
\includegraphics[width=0.65\textwidth]{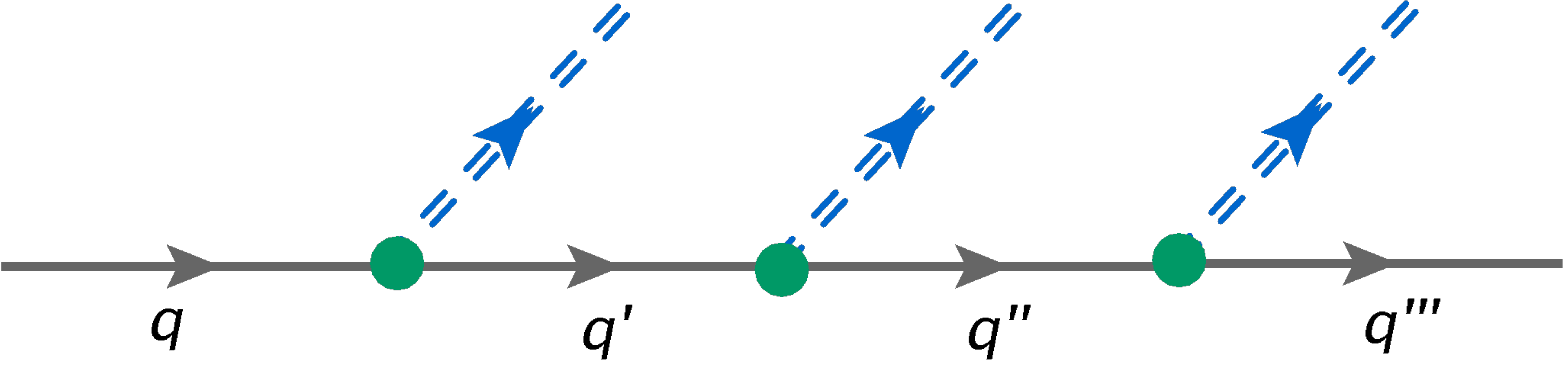}
\caption{Quark cascade.}
\label{PLOT_QUARK_CASCADE}
\end{figure*}

 The recent development of the NJL-Jet model \cite{Ito:2009zc,Matevosyan:2010hh} has led to an ever more sophisticated model for calculating quark fragmentation functions in an effective chiral quark theory. Here the quark fragmentation is modeled as a quark-cascade process, depicted in Fig.~\ref{PLOT_QUARK_CASCADE}, and the multiplicative ansatz of Ref.~\cite{Field:1977fa} was used to derive a set of coupled integral equations for the fragmentation functions. The advantages of the NJL-Jet model are the absence of ad-hoc parameters included, i.e. all the NJL parameters are fixed independent of any experimental data on fragmentation functions. Moreover, the momentum and isospin sum rules for the solutions of fragmentation functions are naturally satisfied. On the other hand, the model assumes that the initial quark has infinite momentum (Bjorken limit) and emits an infinite number of hadrons, so the probabilities scale with the fraction of the light-cone momentum left in the quark cascade, while in medium-energy experiments only a few hadrons are emitted per struck quark. Another problem appears as more and more possible emission channels are included in model, as the numerical task of solving the corresponding integral equations becomes challenging. Lastly, the inclusion of the transverse momentum dependence of the fragmentation functions seems unachievable in the integral equation approach. 
 
 Here we propose to use Monte Carlo method to solve for the fragmentation functions within the quark-cascade model. Our goal is to demonstrate the viability of the approach of replacing the integral equations by providing very similar order of precision in determining the fragmentation functions over the entire range of the light-cone momentum fraction $z$ of the initial quark carried by the emitted hadrons. 
 \section{Monte-Carlo Approach and the Asymptotic Equivalence to the Integral Equation Method}
 \label{SEC_MC_SIMS}
 
 The NJL-Jet model (\cite{Ito:2009zc,Matevosyan:2010hh}) relied on a set of coupled integral equations to solve for the fragmentation functions:
\begin{equation}
\label{EQ_FRAG_PROB}
D^{h}_{q}(z)dz=\hat{d}^{h}_{q}(z)dz+\sum_{Q}\int^{1}_{z}\hat{d}^{Q}_{q}(y) dy  \; D^{h}_{Q}\left(\frac{z}{y}\right) \frac{dz}{y},
\end{equation}
where we assumed that the quark has infinite momentum and produces an infinite number of hadrons. Here $D^h_q(z)$ denotes the fragmentation function of quark $q$ to hadron $h$ carrying light-cone momentum fraction $z$, $\hat{d}^{h}_{q}(z)$ is the "elementary" fragmentation function of quark $q$ emitting hadron $h$ at each vertex  in the quark-cascade and the sum on the right hand side is over all active flavors of quarks included in the model.
 
  Here we propose to calculate the fragmentation functions using Monte-Carlo (MC) simulations akin to the method described in the Ref.~\cite{Ritter:1979mk} using the probabilistic interpretation of the fragmentation functions: $D_q^h(z)$  is the probability to emit a hadron $h$ carrying the light-cone momentum fraction $z$ to $z + dz$ of initial quark $q$ in a quark-jet picture. The quark goes through a cascade of hadron emissions, where at every emission vertex we choose the type of emitted hadron $h$ and its light-cone momentum fraction $z$ (of the fragmenting quark) by randomly sampling the corresponding probabilities of the elementary fragmentations, $\hat{d}_q^h(z)$ that are calculated within the NJL model (in general these can be calculated in any effective quark model). We keep track of the flavor and the light-cone momentum fraction of the initial quark left to the remaining  quark, also recording the type and light-cone momentum fraction of the initial quark transferred to the emitted hadron. We stop the fragmentation chain after the quark has emitted a predefined number of hadrons, $N_{Links}$ (Alternatively, we can stop the chain after the remnant quark in the cascade  has less than a given fraction of the initial quark's light-cone momentum, $z_{Min}$). We repeat the calculation $N_{Sims}$ times with the same initial quark flavor, $q$ until we have sufficient statistics for the emitted hadrons. We extract the fragmentation functions by calculating the average number of hadrons of type $h$ with light-cone momentum fraction $z$ to $z + \Delta z$, $\left<N_q^h(z, z+ \Delta z) \right>$ and expressing them in terms of fragmentation functions:

\begin{eqnarray}
\label{EQ_FRAG_MC}
D_q^h(z) \Delta z = \left< N_q^h(z, z+ \Delta z) \right> \equiv  \frac{ \sum_{N_{Sims}} N_q^h(z, z+ \Delta z) } { N_{Sims} }
\end{eqnarray}

\begin{figure}[phtb]
\centering 
\includegraphics[width=0.8\textwidth]{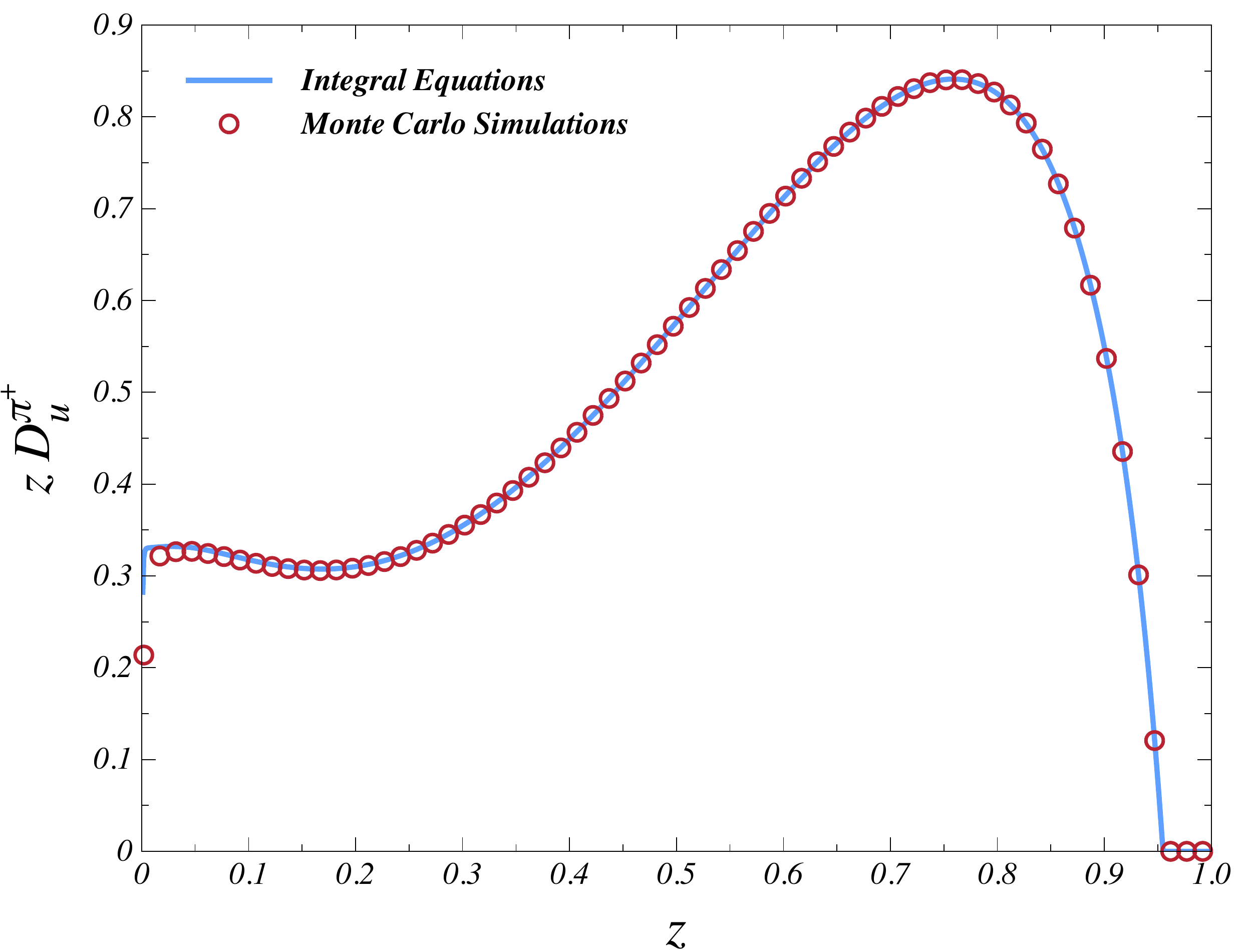}
\caption{Comparison of  the solutions for quark fragmentation function $D_u^{\pi^+}(z) $ in NJL-Jet model with only non-strange pseudoscalar mesons calculated from integral equations Eq.~(\ref{EQ_FRAG_PROB}) and MC simulation.}
\label{PLOT_MC_INT_COMP}
\end{figure}

 From the construction it is obvious that the fragmentation functions calculated using the integral Eq.~(\ref{EQ_FRAG_PROB}) should be equivalent to those calculated using the MC method in the limit $N_{Links} \to \infty$ and $N_{Sims} \to \infty$. The plots in Fig.~\ref{PLOT_MC_INT_COMP} show that the solutions for fragmentation functions from both methods are indeed equivalent with a large enough number of emitted hadrons within statistical errors.

\begin{figure}[phtb]
\centering 
\includegraphics[width=0.8\textwidth]{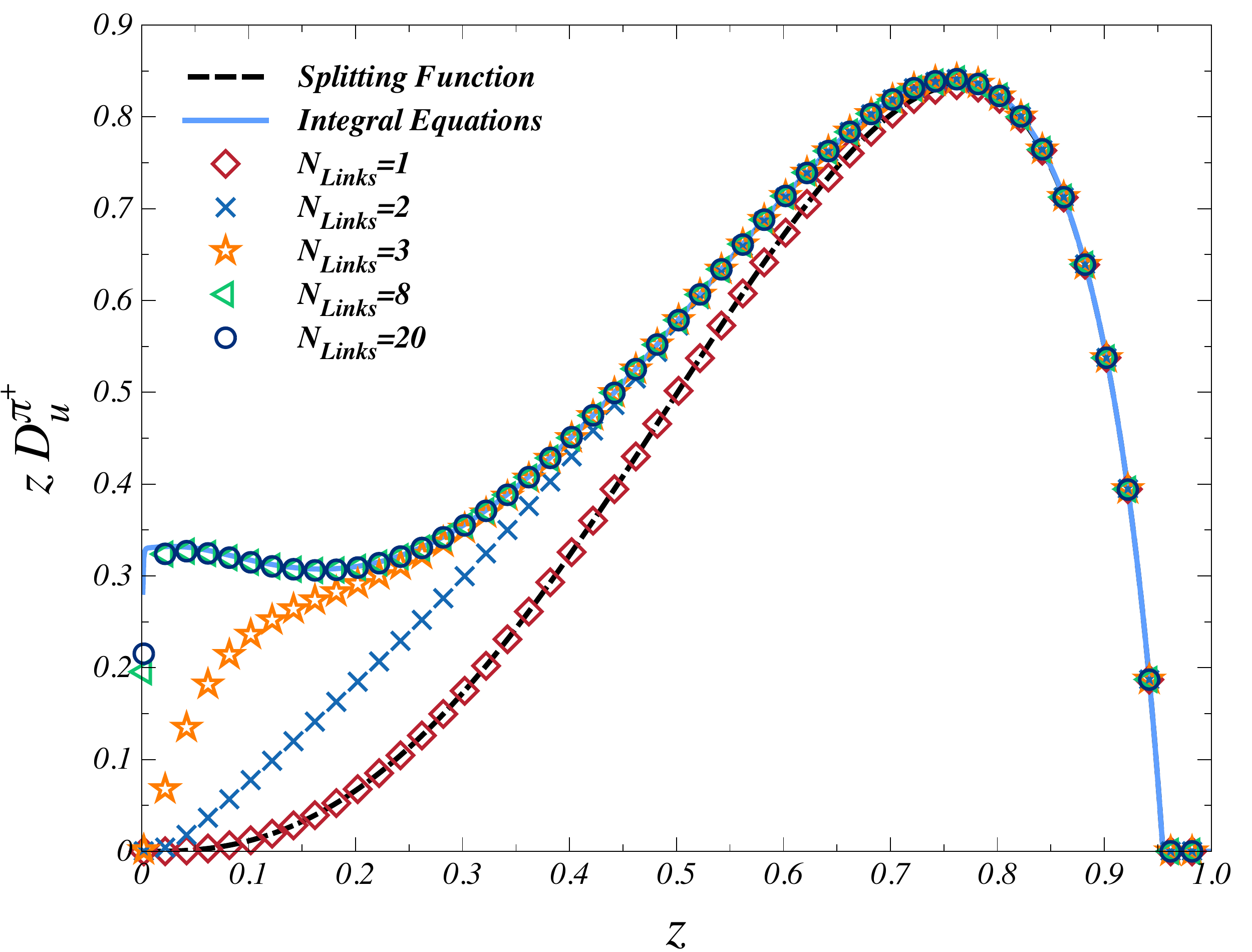}
\caption{The dependence of the solutions for $z D_u^{\pi^+}$ on $N_{Links}$.}
\label{PLOT_FRAG_VS_NLINKS}
\end{figure}

  MC also allows us to study the dependence of the resulting fragmentation functions on the number of hadrons emitted by the quark in the cascade, which could well be relevant to many medium-energy experiments. The plot in Fig.~\ref{PLOT_FRAG_VS_NLINKS} show that the solution for $z D_u^{\pi^+}(z) $ with $N_{Links}=1$ (equivalent to the elementary fragmentation function $z \hat{d}_u^{\pi^+}(z)$) is peaked at  $z \sim 0.8$. As we increase the number of emitted hadrons, the solution increases in the low $z$ region due to the hadrons emitted further in the quark jet, where the fragmenting quark typically has a small fraction of the initial quark's light-cone momentum. We can readily see that the solutions saturate after including only a few emitted hadrons, where there is virtually no difference between solutions with $N_{Links} = 8$ and $N_{Links} = 20$, and the discrepancy with the solution of the integral equations only occurs at vanishingly small values of $z$. Thus we can reliably use the solutions of MC simulations with $N_{Links}\geq 8$.  
  
\section{Final Remarks}

 In the current article we demonstrated that the Monte Carlo approach to calculating the fragmentation functions in NJL-Jet framework is a powerful and reliable method. We reproduced the fragmentation functions calculated as solutions of the previously employed integral equations, where only the light quarks and pions were included. Moreover, we showed that the MC approach allows for the flexibility to surpass the model limitations necessary in formulating the integral equations, i.e. in the future MC studies we can assume an initial quark carrying only a finite momentum, thus emitting a finite number of hadrons. 
 
  A further advantage of the MC approach is in reducing the numerical task in solving for the fragmentation functions when including many more channels for emitted hadrons. Here solving the integral equations requires inverting larger and larger matrix, while the MC procedure can be drastically sped up by trivially parallelizing the task and  solving simultaneously on computer clusters.
  
  It is clear that for a more complete description of the quark fragmentation both vector meson and nucleon anti-nucleon channels, as well as the strong decays of the produced resonances need to be included in the calculations. This is because as the high $z$ region of the fragmentation functions are dominated by ``few-step'' transitions where the availability of the additional fragmentation channels might have a noticeable effect. The future development of the NJL-Jet model would also allow access to the transverse momentum distribution of the produced hadrons, thus becoming relevant for the analysis of a large variety of semi-inclusive data.  The MC approach provides a strong platform for implementing these and other possible extensions of the NJL-Jet model that would allow for a much more detailed description of the physical picture.
\section*{Acknowledgements}
 
This work was supported by the Australian Research Council through the grant of an Australian Laureate Fellowship (A.W. T.) and by a Subsidy for Activating Educational Institutions from the Department of Physics, Tokai University.
  
\bibliographystyle{apsrev}
\bibliography{../Bibliography/fragment}

\begin{thebibliography}{16}
\expandafter\ifx\csname natexlab\endcsname\relax\def\natexlab#1{#1}\fi
\expandafter\ifx\csname bibnamefont\endcsname\relax
  \def\bibnamefont#1{#1}\fi
\expandafter\ifx\csname bibfnamefont\endcsname\relax
  \def\bibfnamefont#1{#1}\fi
\expandafter\ifx\csname citenamefont\endcsname\relax
  \def\citenamefont#1{#1}\fi
\expandafter\ifx\csname url\endcsname\relax
  \def\url#1{\texttt{#1}}\fi
\expandafter\ifx\csname urlprefix\endcsname\relax\def\urlprefix{URL }\fi
\providecommand{\bibinfo}[2]{#2}
\providecommand{\eprint}[2][]{\url{#2}}

\bibitem[{\citenamefont{Hirai et~al.}(2007)\citenamefont{Hirai, Kumano, Nagai,
  and Sudoh}}]{Hirai:2007cx}
\bibinfo{author}{\bibfnamefont{M.}~\bibnamefont{Hirai}},
  \bibinfo{author}{\bibfnamefont{S.}~\bibnamefont{Kumano}},
  \bibinfo{author}{\bibfnamefont{T.~H.} \bibnamefont{Nagai}}, \bibnamefont{and}
  \bibinfo{author}{\bibfnamefont{K.}~\bibnamefont{Sudoh}},
  \bibinfo{journal}{Phys. Rev.} \textbf{\bibinfo{volume}{D75}},
  \bibinfo{pages}{094009} (\bibinfo{year}{2007}), \eprint{hep-ph/0702250}.

\bibitem[{\citenamefont{de~Florian et~al.}(2007)\citenamefont{de~Florian,
  Sassot, and Stratmann}}]{deFlorian:2007aj}
\bibinfo{author}{\bibfnamefont{D.}~\bibnamefont{de~Florian}},
  \bibinfo{author}{\bibfnamefont{R.}~\bibnamefont{Sassot}}, \bibnamefont{and}
  \bibinfo{author}{\bibfnamefont{M.}~\bibnamefont{Stratmann}},
  \bibinfo{journal}{Phys. Rev.} \textbf{\bibinfo{volume}{D75}},
  \bibinfo{pages}{114010} (\bibinfo{year}{2007}), \eprint{hep-ph/0703242}.

\bibitem[{\citenamefont{Field and Feynman}(1978)}]{Field:1977fa}
\bibinfo{author}{\bibfnamefont{R.~D.} \bibnamefont{Field}} \bibnamefont{and}
  \bibinfo{author}{\bibfnamefont{R.~P.} \bibnamefont{Feynman}},
  \bibinfo{journal}{Nucl. Phys.} \textbf{\bibinfo{volume}{B136}},
  \bibinfo{pages}{1} (\bibinfo{year}{1978}).

\bibitem[{\citenamefont{Altarelli et~al.}(1979)\citenamefont{Altarelli, Ellis,
  Martinelli, and Pi}}]{Altarelli:1979kv}
\bibinfo{author}{\bibfnamefont{G.}~\bibnamefont{Altarelli}},
  \bibinfo{author}{\bibfnamefont{R.~K.} \bibnamefont{Ellis}},
  \bibinfo{author}{\bibfnamefont{G.}~\bibnamefont{Martinelli}},
  \bibnamefont{and} \bibinfo{author}{\bibfnamefont{S.-Y.} \bibnamefont{Pi}},
  \bibinfo{journal}{Nucl. Phys.} \textbf{\bibinfo{volume}{B160}},
  \bibinfo{pages}{301} (\bibinfo{year}{1979}).

\bibitem[{\citenamefont{Collins and Soper}(1982)}]{Collins:1981uw}
\bibinfo{author}{\bibfnamefont{J.~C.} \bibnamefont{Collins}} \bibnamefont{and}
  \bibinfo{author}{\bibfnamefont{D.~E.} \bibnamefont{Soper}},
  \bibinfo{journal}{Nucl. Phys.} \textbf{\bibinfo{volume}{B194}},
  \bibinfo{pages}{445} (\bibinfo{year}{1982}).

\bibitem[{\citenamefont{Jaffe}(1996)}]{Jaffe:1996zw}
\bibinfo{author}{\bibfnamefont{R.~L.} \bibnamefont{Jaffe}}
  (\bibinfo{year}{1996}), \eprint{hep-ph/9602236}.

\bibitem[{\citenamefont{Ellis et~al.}(1996)\citenamefont{Ellis, Stirling, and
  Webber}}]{Ellis:1991qj}
\bibinfo{author}{\bibfnamefont{R.~K.} \bibnamefont{Ellis}},
  \bibinfo{author}{\bibfnamefont{W.~J.} \bibnamefont{Stirling}},
  \bibnamefont{and} \bibinfo{author}{\bibfnamefont{B.~R.}
  \bibnamefont{Webber}}, \bibinfo{journal}{Camb. Monogr. Part. Phys. Nucl.
  Phys. Cosmol.} \textbf{\bibinfo{volume}{8}}, \bibinfo{pages}{1}
  (\bibinfo{year}{1996}).

\bibitem[{\citenamefont{Barone et~al.}(2002)\citenamefont{Barone, Drago, and
  Ratcliffe}}]{Barone:2001sp}
\bibinfo{author}{\bibfnamefont{V.}~\bibnamefont{Barone}},
  \bibinfo{author}{\bibfnamefont{A.}~\bibnamefont{Drago}}, \bibnamefont{and}
  \bibinfo{author}{\bibfnamefont{P.~G.} \bibnamefont{Ratcliffe}},
  \bibinfo{journal}{Phys. Rept.} \textbf{\bibinfo{volume}{359}},
  \bibinfo{pages}{1} (\bibinfo{year}{2002}), \eprint{hep-ph/0104283}.

\bibitem[{\citenamefont{Martin et~al.}(2004)\citenamefont{Martin, Roberts,
  Stirling, and Thorne}}]{Martin:2003sk}
\bibinfo{author}{\bibfnamefont{A.~D.} \bibnamefont{Martin}},
  \bibinfo{author}{\bibfnamefont{R.~G.} \bibnamefont{Roberts}},
  \bibinfo{author}{\bibfnamefont{W.~J.} \bibnamefont{Stirling}},
  \bibnamefont{and} \bibinfo{author}{\bibfnamefont{R.~S.}
  \bibnamefont{Thorne}}, \bibinfo{journal}{Eur. Phys. J.}
  \textbf{\bibinfo{volume}{C35}}, \bibinfo{pages}{325} (\bibinfo{year}{2004}),
  \eprint{hep-ph/0308087}.

\bibitem[{\citenamefont{Ralston and Soper}(1979)}]{Ralston:1979ys}
\bibinfo{author}{\bibfnamefont{J.~P.} \bibnamefont{Ralston}} \bibnamefont{and}
  \bibinfo{author}{\bibfnamefont{D.~E.} \bibnamefont{Soper}},
  \bibinfo{journal}{Nucl. Phys.} \textbf{\bibinfo{volume}{B152}},
  \bibinfo{pages}{109} (\bibinfo{year}{1979}).

\bibitem[{\citenamefont{Sivers}(1990)}]{Sivers:1989cc}
\bibinfo{author}{\bibfnamefont{D.~W.} \bibnamefont{Sivers}},
  \bibinfo{journal}{Phys. Rev.} \textbf{\bibinfo{volume}{D41}},
  \bibinfo{pages}{83} (\bibinfo{year}{1990}).

\bibitem[{\citenamefont{Boer et~al.}(2003)\citenamefont{Boer, Mulders, and
  Pijlman}}]{Boer:2003cm}
\bibinfo{author}{\bibfnamefont{D.}~\bibnamefont{Boer}},
  \bibinfo{author}{\bibfnamefont{P.~J.} \bibnamefont{Mulders}},
  \bibnamefont{and} \bibinfo{author}{\bibfnamefont{F.}~\bibnamefont{Pijlman}},
  \bibinfo{journal}{Nucl. Phys.} \textbf{\bibinfo{volume}{B667}},
  \bibinfo{pages}{201} (\bibinfo{year}{2003}), \eprint{hep-ph/0303034}.

\bibitem[{\citenamefont{Ito et~al.}(2009)\citenamefont{Ito, Bentz, Cloet,
  Thomas, and Yazaki}}]{Ito:2009zc}
\bibinfo{author}{\bibfnamefont{T.}~\bibnamefont{Ito}},
  \bibinfo{author}{\bibfnamefont{W.}~\bibnamefont{Bentz}},
  \bibinfo{author}{\bibfnamefont{I.~C.} \bibnamefont{Cloet}},
  \bibinfo{author}{\bibfnamefont{A.~W.} \bibnamefont{Thomas}},
  \bibnamefont{and} \bibinfo{author}{\bibfnamefont{K.}~\bibnamefont{Yazaki}},
  \bibinfo{journal}{Phys. Rev.} \textbf{\bibinfo{volume}{D80}},
  \bibinfo{pages}{074008} (\bibinfo{year}{2009}), \eprint{0906.5362}.

\bibitem[{\citenamefont{Matevosyan et~al.}(2010)\citenamefont{Matevosyan,
  Thomas, and Bentz}}]{Matevosyan:2010hh}
\bibinfo{author}{\bibfnamefont{H.~H.} \bibnamefont{Matevosyan}},
  \bibinfo{author}{\bibfnamefont{A.~W.} \bibnamefont{Thomas}},
  \bibnamefont{and} \bibinfo{author}{\bibfnamefont{W.}~\bibnamefont{Bentz}}
  (\bibinfo{year}{2010}), \eprint{1011.1052}.

\bibitem[{\citenamefont{Ritter and Ranft}(1980)}]{Ritter:1979mk}
\bibinfo{author}{\bibfnamefont{S.}~\bibnamefont{Ritter}} \bibnamefont{and}
  \bibinfo{author}{\bibfnamefont{J.}~\bibnamefont{Ranft}},
  \bibinfo{journal}{Acta Phys.Polon.} \textbf{\bibinfo{volume}{B11}},
  \bibinfo{pages}{259} (\bibinfo{year}{1980}).

\bibitem[{\citenamefont{Field and Feynman}(1977)}]{Field:1976ve}
\bibinfo{author}{\bibfnamefont{R.~D.} \bibnamefont{Field}} \bibnamefont{and}
  \bibinfo{author}{\bibfnamefont{R.~P.} \bibnamefont{Feynman}},
  \bibinfo{journal}{Phys. Rev.} \textbf{\bibinfo{volume}{D15}},
  \bibinfo{pages}{2590} (\bibinfo{year}{1977}).

\end{thebibliography}

\end{document}